\documentclass[12pt]{iopart}
\usepackage{graphics}
\usepackage{amssymb}
\begin{document}
\jl{3}
\eqnobysec

\title{Wedge filling, cone filling and the strong-fluctuation regime}

\author{A. O. Parry, A. J. Wood and C.Rasc\'{o}n }
\address{Department of Mathematics, Imperial College 180 Queen's Gate, London SW7 2BZ, United Kingdom}

\begin{abstract}
Interfacial fluctuation effects occuring at wedge and cone filling
transitions are investigated and shown to exhibit very different
characteristics. For both geometries we show how the conditions for
observing critical (continuous) filling are much less restrictive than
for critical wetting, which is known to require fine tuning of the
Hamaker constants. Wedge filling is critical if the wetting binding
potential does not exhibit a local maximum, whilst conic filling is
critical if the integrated strength of the potential is attractive. This
latter scenario is particularly encouraging for future experimental
studies. 
 
Using mean-field and effective Hamiltonian approaches, which allow for
breather-mode fluctuations which translate the interface up and down the
sides of the confining geometry, we are able to completely classify the
possible critical behaviour (for purely thermal disorder). For the three
dimensional wedge, the interfacial fluctuations are very strong and
characterised by a universal roughness critical exponent $\nu_{\perp} =1/4$
independent of the range of the forces. For the physical dimensions
$d=2$ and $d=3$, we show that the influence of the cone geometry on the
fluctuations at critical filling is to mimic the analogous interfacial
behaviour occuring at critical wetting in the strong-fluctuation
regime. In particular, for $d=3$ and for quite arbitary choices of
intermolecular potential, the filling height and roughness show the same
critical properties as those predicted for three dimensional critical
wetting with short-ranged forces in the large wetting parameter
($\omega>2$) regime.  

\end{abstract}

\maketitle

\section{Introduction}
Following the seminal studies of Cahn \cite{Cahn} and Ebner and Saam
\cite{EandS}, two fundamental questions concerning the nature of the
wetting transition have drawn an enormous amount of attention
\cite{General, Forgacs}: Firstly, is the transition first-order or
continuous? In particular, how does the order depend on the range and
balance of the competing intermolecular forces? Secondly, for continuous
(critical) wetting transitions, what role do the interfacial fluctuations
of the unbinding fluid interface play in determining the critical
exponents characterising the diverging lengthscales?

The answers to both these questions reveal a host of reasons why the
laboratory observation of fluctuation effects at critical wetting is
beset with difficulties. In brief, the well developed fluctuation theory
of wetting concludes \cite{General, Forgacs} that three dimensional
critical wetting requires a fine tuning of the effective Hamaker
constants appearing in the binding potential. This sensitivity means that
critical wetting is a rather rare phenomena of which only very few
examples are known for fluid-fluid interfaces
\cite{Ragil, Ross}.  Secondly, for systems with long-ranged forces, the
critical exponents at critical wetting are mean-field-like (and depend on
the range of the forces) and the interfacial roughness induced by
fluctuations is very small. Even for short-ranged forces, where
renormalization group (RG) theory based on effective interfacial models
famously predicts dramatic fluctuation induced effects \cite{RG},  no
appreciable deviations from simple mean-field like behaviour have been
observed in either Ising model simulation \cite{Simulation} or recent
experimental studies \cite{Ross}.  At the very least, this indicates that
the asymptotic critical regime for critical wetting is much smaller
than initially expected \cite{3DParry}. The conclusion inevitably forced
upon us appears to be that fluctuation effects at wetting transitions
are of negligible practical importance.

The purpose of the present article is to compare and contrast these
features of critical wetting transitions with the properties of
continuous (critical) filling transitions occuring in both wedge and
cone shaped geometries. The thermodynamics of the filling transition has
been discussed independently by a number of authors \cite{Finn,  Pomeau, 
Hauge} but only recently have the analogues of the above questions been
addressed \cite{MFT, ourPRL1, ourJphysCM, ourPRL2, RandN}. Here we present full
details of a novel interfacial theory of wedge filling and also extend
this approach to study fluctuation effects occuring at the filling
transition of a fluid adsorbed in a cone shaped geometry. These turn out
to be completely different to those predicted to occur in the wedge. In
particular we wish to draw attention to two striking features of
conic filling which have not been reported before. First, that the conditions
for critical conic filling, even more than conditions for critical wedge
filling, do not rely on any fine tuning of the binding potential
structure and should be observable in cone geometries made from
substrates that exhibit even strong first-order wetting. Secondly we
point out what appears to be an unexpected connection between the
{\it geometry-induced} aspects of critical conic filling and the
theory of the universal {\it strong-fluctuation} (SFL) regime of critical wetting
\cite{Forgacs,  LandF}.  Specifically, from analysis of effective
interfacial Hamiltonian models, we show that for the physically relevant
dimensions $d=2$ and $d=3$, and also for almost all choices of the
intermolecular interactions, the influence of the conic geometry on the
asymptotic critical behaviour and the detailed scaling structure of the interfacial height probability distribution function (PDF) as one approaches a critical cone
filling transition is to mimic the way in which interfacial fluctuations 
determine the analogous exponents and the PDF for critical wetting transitions 
for planar wall-fluid (or fluid-fluid) interfaces belonging to the SFL scaling regime. This remarkable connection between conic filling and
wetting and has already been noted in $d = 2$ from the explicit results of exact
transfer matrix and replica trick studies of two-dimensional filling
transitions (where, of course, the cone is analogous to a wedge) in both
pure and impure bulk systems \cite{ourPRL1,  ourJphysCM}. Extending this
theory our analysis
shows that the fluctuation properties of three-dimensional conic filling are
quite distinct from those occuring at three-dimensional wedge filling
\cite{ourPRL2} and are instead related to some of the most dramatic
predictions of the theory of wetting for planar wall-fluid
interfaces. Thus in $d=3$ the divergence of the equilibrium filling layer
thickness and roughness at cone filling are universal and the same as
the renormalization group predictions of Brezin, Halperin and Leibler \cite{RG}
for critical wetting with short-ranged forces in the large wetting
parameter ($\omega  >2$) regime. 

Our article is arranged as follows. In the next section we briefly recall
the well known fluctuation theory predictions for critical wetting detailing
the critical behaviour predicted to occur in the SFL regime in $d=2$ and
$d=3$.  In \S 3 we present the main methods and results of the
fluctuation theory of three dimensional wedge filling emphasising both
the conditions for critical filling and the predictions for fluctuation
dominated behaviour. A brief summary of these results has been given
before \cite{ourPRL2}.  We finish this section by turning to the special
case of filling in $d=2$.  The role of a breather-like soft mode which
increases the width of the interface as it translates up and down the
wedge is shown to be directly responsible for the scaling behaviour of
the PDF.  In \S 4 we discuss three dimensional cone filling, at mean-field
level and beyond, predicting universal geometry-dominated critical behaviour
characterised by an appreciable interfacial roughness and critical
exponents identical to those predicted for the planar wetting SFL regime. We argue that this equivalence of the PDF's at cone filling
and SFL wetting is special to $d=2$ and $d=3$ only. We finish our article
with a summary of our main results. 

\section{Fluctuation regimes at critical wetting. }

Consider the preferential adsorption of a thick liquid film at the interface 
between a {planar} wall and a bulk vapour phase at chemical potential
$\mu$ and a sub-critical temperature $T$. We suppose that at bulk two phase coexistence, 
corresponding to chemical potential $\mu = \mu_{\mbox{sat}}$, the wall-vapour
interface is completely wet by the liquid phase for temperatures $T >
T_{\pi}$.  Here we are interested in the values of the critical exponents
describing the continuous divergence of the mean wetting film thickness
$l_{\pi} \sim t'^{-\beta_s}$,  roughness $\xi_{\perp} \sim t'^{-\nu_{\perp}}$ and transverse correlation length
$\xi_{\parallel} \sim t'^{- \nu_{\parallel}}$ as $t' = (T_{\pi} -
T)/T_{\pi} \rightarrow 0$. The
subscript $\pi$ is used to emphasise that these quantities are pertinent to
planar wetting transitions. We denote the reduced temperature $t'$ to
avoid confusion with the filling reduced temperature (q.v.). The
standard theory of interfacial fluctuation effects at the wetting
transition is based on analysis of the effective interfacial Hamiltonian \cite{General, Forgacs} 
\begin{equation} 
\label{effham}
H[l] = \int d {\bf x} \left\{ \frac{\Sigma}{2} \left(\nabla l  \right)^2 + W(l) \right\}
\end{equation}
where $l$ is a collective co-ordinate describing the local height of the
interface above the wall,  $\Sigma$ is the surface stiffness (tension) of the 
unbinding liquid-vapour interface and $W(l)$ is the usual binding
potential which accounts for the direct influence of the molecular
forces. Note that throughout this article we shall suppose that the factor
$\beta (\equiv 1/k_B T)$ is absorbed into all the effective Hamiltonians, stiffnesses and
potentials etc. Mean-field (MF) studies indicate that for critical
wetting (with long-ranged forces) $W(l)$ is necessarily of the form 
\begin{equation}
W(l) = -\frac{a}{l^p} + \frac{b}{l^q} + \dots ; l > 0 
\label{BPot} 
\end{equation}
where $a \propto (T_{\pi}^{MF} - T)$ and $b > 0$ are effective Hamaker constants. Here the exponents $p$
and $q$ are specific to the particular ranges of the intermolecular forces. In 
$d=3$ and for van der Waals forces the appropriate values are $p = 2$
and $q = 3$.  For fixed dimension $d<3$ the scaling theory of Lipowsky and
Fisher \cite{LandF} (which is supported by transfer-matrix and
approximate non-linear renormalization group (RG) studies) predicts that the critical behaviour
falls into three generic fluctuation classes labelled mean-field
(MF), weak-fluctuation (WFL) and strong-fluctuation (SFL) scaling
regimes. These regimes arise due to the interplay of the direct binding
potential with the entropic repulsion originating from the interfacial wandering.  We
briefly recall the scaling argument here since the same ideas
generalise to filling transitions in wedge geometries. First recall that
for interfaces, the roughness $\xi_{\perp}$ and the transverse
correlation length $\xi_{\parallel}$ are related via the wandering
exponent $\xi_{\perp} \sim \xi_{\parallel}^{\zeta_{\pi}}$. We write $\zeta_{\pi}$ keeping the
sub-script to avoid
confusion with different wandering exponents that emerge for filling in
wedges and cones. For thermal forces and $d<3$ it is well known that
\begin{equation}
\zeta_{\pi} = \frac{3 - d}{2}
\label{Wetzeta}
\end{equation}
with $\xi_{\perp}$ finite for $d > 3$ and $\xi_{\perp} \sim \sqrt{\ln} \xi_{\parallel}$ for the marginal
case. Thus we are led rather naturally \cite{LandF} to deduce that the
bending energy term in \eref{effham} and also the entropic effect of
interface collisions with the wall gives rise to a fluctuation
contribution to the effective potential of the form $l^{-\tau}$ with
$\tau = 2(d-1)/(3 - d)$ for $d < 3$. The competition between these three terms gives rise to the
three fluctuation regimes mentioned above. In the MF and WFL regimes the
leading term in \ref{BPot} is still dominant and the wetting transition
temperature is unchanged by fluctuations. The critical exponents are non-universal and are given by \cite{LandF}
\begin{equation}
\beta_s^{MF} = \frac{1}{q-p}, \quad \nu_{\parallel}^{MF} =
\frac{q+2}{2(q-p)},\quad \nu^{MF}_{\perp} = \zeta_{\pi} \nu_{\parallel}^{MF} 
\label{WetCosMF}
\end{equation}
\begin{equation}
\beta_s^{WF} = \frac{1}{\tau-p}, \quad\nu_{\parallel}^{WF} =
\frac{\tau+2}{2(\tau-p)}, \quad
\nu^{WF}_{\perp} = \zeta_{\pi} \nu_{\parallel}^{WF}
\label{WetCosWF}
\end{equation}

In the SFL, corresponding to $p < \tau$, and representing the critical
behaviour in systems with short-ranged forces, the transition
temperature is lowered by fluctuations and the critical exponents are
universal. In $d=2$ the critical exponents and the interfacial
PDF, $P_{\pi} (l)$, are very
well known (see for example \cite{Abraham,Burkhardt} and also references
in \cite{General, Forgacs}) and given by
\begin{equation}
\beta_s^{SF} = 1, \quad \nu_{\perp}^{SF} = 1, \quad \nu^{SF}_{\parallel} = 2
\mbox{ and } P_{\pi} (l) = \frac{1}{l_{\pi}} e^{-\frac{l}{l_{\pi}}}
\label{2WetCos} 
\end{equation}

Notice that in both the SFL and WFL the interfacial fluctuations are
very strong and $l_{\pi} \sim \xi_{\perp}$ for $d < 3$. The unusual
bifurcation mechanism controlling the fixed point behaviour as $d
\rightarrow 3$ means that critical wetting in three-dimensions systems is
rather different in the SFL regime, though for systems with long-ranged
forces MF theory remains valid. Fluctuation effects are only predicted
to occur for systems with short-ranged forces and are traditionally and
most simply described by a bare binding potential of the form
\cite{General, Forgacs, RG, 3DParry}
\begin{equation}
W(l) = -a e^{-\kappa l} + b e^{-2 \kappa l} + \dots; l > 0 
\label{BPotS} 
\end{equation}
together with the hard-wall restriction $W(l) = \infty$ for $l < 0$. Here
$\kappa$ denotes the inverse bulk correlation length of the adsorbed
bulk phase whilst the Hamaker constants $a$ and $b$ have the same
interpretation as for long-ranged forces discussed above. The mean-field
predictions for short-ranged critical wetting are therefore $\beta_s = 0$
(ln) and $\nu_{\parallel} = 1$. The fluctuation theory of critical
wetting with short-ranged forces was famously investigated by Brezin,
Halperin and Leiblier \cite{RG} using a linearised RG scheme and shows
that the critical behaviour is strongly non-universal and falls into
three regimes (labelled I - III) depending on the value of the
dimensionless wetting parameter $\omega$ where 
\begin{equation}
\omega = \frac{k_B T_{\pi} \kappa^2}{4 \pi \Sigma}
\end{equation}
The three regimes 
correspond to the ranges $\omega < 1/2$, $1/2 < \omega < 2$ and $\omega > 2$ respectively. Usually attention focuses on the value of the transverse correlation length critical exponent which shows the strongest non-universal critical behaviour:
\begin{equation}
\mbox{(I)} \quad \nu_{\parallel} = \frac{1}{1 - \omega}, \quad
\mbox{(II)} \quad\nu_{\parallel} = \frac{1}{(\sqrt{2} -
\sqrt{\omega})^2},\quad \mbox{(III)} \quad \nu_{\parallel} = \infty
\label{SFLregimes}
\end{equation}
where the last case corresponds to an essential singularity. In all three
regimes the roughness $\xi_{\perp} \sim \sqrt{l_{\pi}}$ so that
interfacial wandering has a much more pronounced influence on the
equilibrium distribution of matter in the wetting film compared to
systems with long-ranged forces. In regimes (I) and (II) the wetting
temperature is unchanged from its mean-field value and the film
thickness still grows logarithmically but with a different amplitude to
mean-field theory. The most dramatic effect of interfacial wandering is
manifest in the regime (III) which bears all the hallmarks of the SFL
regime.Here the wetting temperature is depressed by the fluctuations, and the
critical exponents are given by
\begin{equation}
\mbox{(III)} \quad \beta_s = 1, \quad \nu_{\perp} = \frac{1}{2}
\label{SFL3}
\end{equation}
with an equilibrium PDF which is a simple Gaussian. 
     
\section{Fluctuation Regimes at Critical Wedge Filling.}

\subsection{Three-dimensional wedge filling:  phenomenology.}

   Consider a three-dimensional wedge formed by the junction of two walls at
angles $\pm \alpha$ to the horizontal with the height of the wall above the
plane is described by a height function $z_w (x,y) \equiv \alpha \vert x \vert$. Thus
the wedge has a `V' shaped cross-section in the $x$-direction with the
wedge bottom oriented along the $y$ axis. Following our earlier
discussion of wetting at planar walls we suppose that the wedge is in
contact with a bulk vapour phase at saturation chemical potential $\mu =
\mu_{\mbox{sat}}$. General thermodynamic considerations \cite{Finn,
Pomeau, Hauge} indicate that the wedge is completely filled by liquid
for temperatures greater then the filling temperature $T_F$, which is
specified by the elegant equality
\begin{equation}
\theta_{\pi}(T_F) = \alpha 
\label{fillT}
\end{equation}
 where $\theta_{\pi}(T)$ denotes the temperature dependent contact
angle of the liquid drop at the planar wall-vapour interface. Thus the
influence of the wedge geometry is to lower the temperature at which the
liquid-vapour interface unbinds from the wedge bottom compared to the
planar wall. The filling transition occurring as $t = (T_F -T)/T_F \rightarrow 0$ may
be first-order or critical corresponding to the discontinuous or
continuous divergence of the equilibrium interfacial height
$\langle l_0 \rangle \equiv \langle l(x = 0,y) \rangle$
measured from the bottom of the wedge. Critical filling is also
characterised \cite{ourPRL2} by the divergence of distinct correlation
lengths describing fluctuations in the interfacial height along
($\xi_y$) and across ($\xi_x$) the wedge and also by the roughness
$\xi_\perp$ of the unbinding interface. Note that the
cross-wedge correlation length essentially measures the width of the
flat, filled region of the wedge. The flatness of the filled region
simply reflects the absence of any macroscopic curvature of the meniscus
at two phase coexistence. Recalling our earlier notation we define the wedge wetting critical exponents via
\begin{equation}
\langle l_0 \rangle \sim t^{-\beta_s^{W}}, \quad  \xi_{\perp} \sim t^{-\nu_{\perp}^W},
\quad \xi_{x} \sim t^{-\nu_{x}}, \quad \xi_{y} \sim t^{-\nu_{y}}
\label{FillExps}
\end{equation}
where we have adopted a suitable super-script, $W$ (for wedge), for some of the 
exponents to avoid confusion with those pertinent to critical wetting and also
for cone filling which we will consider later. 

Wedge filling can also be studied using a simple modification of the 
effective interfacial Hamiltonian and for rather open wedges corresponding to
small $\alpha$ (where $\tan \alpha \approx \alpha$) the appropriate model
is \cite{MFT, ourPRL1, ourJphysCM, ourPRL2}
\begin{equation}
H[l] = \int d x \left\{\frac{\Sigma}{2} \left(\nabla l \right)^2 + W(l -
\alpha \vert x \vert) \right\} 
\label{EffHamW} 
\end{equation}
where $l(x,y)$ denotes the height of the interface measured above the
plane. As shown by Rejmer \etal \cite{MFT} this model can be justified
from analysis of a more general drum-head model and with a binding
potential interaction depending on the local normal distance of the
interface to the wall rather than the simple vertical height used in
\eref{EffHamW}. The simplifying approximations implicit in the above model
do not effect any of the universal physics associated with the filling
phase transition. 

\subsection{Mean-field Theory}

At MF level the equilibrium interfacial profile follows from the solution to
the Euler-Lagrange equation found from the minimization of \eref{EffHamW}
\begin{equation}
\Sigma \ddot{l} = W'(l - \alpha \vert x \vert)
\label{E-L}
\end{equation}
which is solved subject to the boundary conditions that $\dot{l} (0)
\equiv dl / dx \big|_0 =0$ and
$l(x) \rightarrow l_{\pi} \pm \alpha x$ as $x \rightarrow \pm \infty$. This equation can be integrated once and leads to simple condition for the mid-point height \cite{MFT, ourPRL2}:
\begin{equation}
\frac{\Sigma \alpha^2}{2} = W(\langle l_0 \rangle) - W(l_{\pi})
\label{E-LCond}
\end{equation}

Note that in the limit $\langle l_0 \rangle \rightarrow \infty$, the RHS $\rightarrow \Sigma \theta_{\pi}^2 / 2$ which follows from
the Youngs equation in the present small
angle limit. The above equation has an elegant graphical
interpretation \cite{MFT} from which one can determine the order of the
filling phase transition occurring as $T \rightarrow T_F$ at fixed
$\alpha$ or equivalently as $\alpha \rightarrow \theta_{\pi}$ at fixed temperature. Thus the MF condition for
first-order/critical wedge filling is that, at the filling temperature
$T_F$, the binding potential does/does not possess a local maximum
separating the minimum at $l = l_{\pi}$ from the extremum at
infinity. Thus wedges made from walls that exhibit critical wetting,
show critical filling whilst wedges made from walls that exhibit
first-order wetting show both first order and critical filling depending
on whether the filling temperature is above or below the spinodal
temperature $T_s$ at which the first-order wetting binding potential develops
a local maximum. Note that since the filling temperature can be lowered
almost arbitarily by increasing the wedge angle $\alpha$ it follows that
the wedge will typically show first-order filling if $\alpha$ is small
and critical filling for larger values of $\alpha$. The tricritical
value of at which the order of the filling transition changes character
is given by $\alpha^* = \theta_{\pi} (T_s)$ the precise value of which depends on how weak the
first-order wetting transition is. Of course this argument presupposes
that a spinodal temperature for the binding potential exists. This is
certainly true for systems with short-ranged forces where it follows
from the Landau theory of wetting. For example in the global
phase diagrams of Nakanishi and Fisher \cite{NandF} the locus of
spinodal temperatures (for different surface fields) corresponds to the
analytic extension of the critical wetting curve. The prediction that wedges made from walls exhibiting
first-order wetting show critical filling can certainly be tested in
Ising model simulation studies. For experimental systems it may be that
the mean-field condition for critical filling cannot be met because the
spinodal temperature $T_s$ is too low. However as we shall see, fluctuation effects lead to a
second mechanism for critical filling beyond the MF condition described
above which may overcome this difficulty. Also, and perhaps of greater
practical importance, we shall show that the mechanism for critical
filling in a cone is rather different to that for a wedge and is much more easily fulfilled. 
  
To continue, from \eref{FillExps} it immediately follows that at a critical filling
transition the MF value of the order-parameter critical exponent is simply determined by the leading order decay of $W(l)$. The exponent $q$ plays no
role and we find $\beta_s^{W} = 1/p$ which is totally different to the
corresponding MF critical exponent $\beta_s \equiv 1/(q-p)$ for critical
wetting. The MF values of the other critical exponents follow from
analysis of the height-height correlation function (with
$\widetilde{y}\!\equiv\!y'\!-\!y$)
\begin{equation}
H(x,x';\widetilde{y})\!\equiv\!\langle\,\delta l(x,y)\, \delta l(x',y')\,\rangle
\label{Hxx'}
\end{equation}
and its Fourier transform       
\begin{equation}
S(x,x';Q)=\int\!d\widetilde{y}\;e^{i\,Q\,\widetilde{y}}\,H(x,x';\widetilde{y})
\label{Sxx'Q}
\end{equation}
which exploits the translational invariance along the wedge. At MF
level, $S(x,x';Q)$ follows from the solution to the Ornstein-Zernike equation
\begin{equation}
\int dx'' C(x,x'';Q) H(x'',x';Q) = \delta (x - x')
\label{OZ}
\end{equation}
where the direct-correlation function is defined by
\begin{equation}
C(x,x';Q) \equiv \frac{\delta^2 H [l]}{\delta l (x) \delta l (x')}
\label{Cxx'Q}
\end{equation}
evaluated at the equilibrium height. Thus we arrive at the differential equation 
\begin{equation}
\left(-\Sigma\partial^{2}_{x}+\Sigma Q^{2}+
W''( l(x)\!-\!\alpha|x|)\right)\!S(x,x';Q)\!=\!\delta(x\!-\!x')
\label{DiffforS}
\end{equation}
To proceed it is best to consider the moment expansion
\begin{equation}
S(x,x';Q) \equiv \sum_{n=0}^{\infty} Q^{2n} S_{2n}(x,x')
\label{Moments}
\end{equation}
and first concentrate on the zeroth moment $S_0 (x, x')$ which describes
the position dependence of correlations in the fluctuations of the
interfacial height across the wedge. This is found by using standard
techniques to be
\begin{eqnarray}
\fl S_0(x,x')=\left(|\dot{l}(x)|-\alpha\right)
\left(|\dot{l}(x')|-\alpha\right) \left\{\frac{1}{2\alpha
W'(l_0)} + \frac{H(xx')}{\Sigma} \int_{0}^{\min(|x|,|x'|)}\!\!
\frac{dx}{(\dot{l}(x)-\alpha)^2}\right\} \nonumber \\
\label{S0xx'}
\end{eqnarray}
\begin{figure}
\begin{center}\resizebox{0.6\textwidth}{!}{%
  \includegraphics{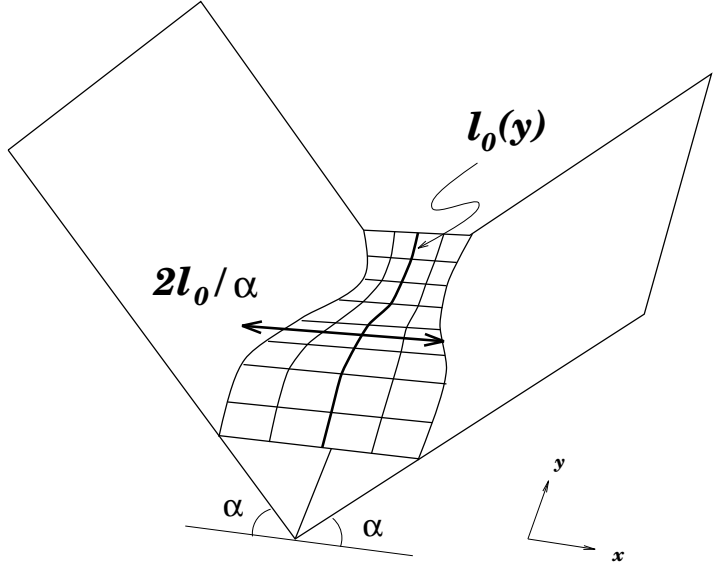}
}
\caption{A depiction of the coherent fluctuations of the interface along
the wedge. The diagram shows the relevant lengths associated with the
derivation of the effective one-dimensional Hamiltonian that describes
such fluctuations}
\end{center}
\label{Fig2}
\end{figure}  
where $H(x)$ denotes the Heaviside step function $(H(x) = 1;x > 0, H(x) =
0; x < 0)$. Thus the structure of correlations across the wedge reflects
the properties of the equilibrium height profile $l(x)$ found from
the solution to the Euler-Lagrange equation (\ref{E-L}), which is
particularly simple. In essence, the interface is flat such that $l(x) \approx
\langle l_0 \rangle$, for $x \lesssim \langle l_0 \rangle / \alpha $, while for
$x \gtrsim \langle l_0 \rangle / \alpha$ the height decays exponentially
quickly to its planar value $l_{\pi}$. Importantly, the lengthscale
which controls the exponential decay is the wetting correlation length
$\xi_{\parallel}$, which remains microscopic at
the filling transition temperature. One consequence of this is that we
can anticipate that the lateral width of the filled region of the wedge
is trivally identified as $\xi_x \sim 2 \langle l_0 \rangle / \alpha$ so
that $\nu_x = \beta_s^{W}$. In this way we conclude that $S_0 (x,x')$ is essentially position independent provided $\vert x
\vert, \vert x' \vert < \xi_x/2$ implying that the whole filled portion of
the wedge fluctuates coherently. Moreover that there is negligible correlation between height fluctuations inside
the wedge with those outside and simlarly between positions on either
side of the filled region. From \eref{S0xx'} one can
iteratively construct the position dependence of the higher moments using the integral equation
\begin{equation}
S_{2n+2} (x,x') = -\Sigma \int S_0 (x,x'') S_{2n} (x'',x') dx''
\label{S2n+2Def}
\end{equation}
which follows from the Ornstein-Zernike equation above. From this it follows
that in the asymptotic critical (scaling) regime $S_{2n} (x,x')$ is
position independent for $\vert x \vert, \vert x' \vert < \xi_x/2$ with  
\begin{equation}
S_{2n+2} (x,x') = \frac{\alpha}{2 W(\langle l_0 \rangle)} \left(
\frac{- \Sigma \langle l_0 \rangle }{W'(\langle l_0 \rangle)}\right)^n
\label{S2n+2Sol}
\end{equation}
and is negligible otherwise. This behaviour is consistent with the simple
Lorentzian structure factor
\begin{equation}
S(x,x';Q) = \frac{S_0 (x,x')}{1 + \xi_y^2 \xi_{\parallel}^2} ; \quad \vert x \vert, \vert x' \vert < \xi_x/2
\label{Sxx'QLor}
\end{equation}
where $S_0 (0,0) \equiv \alpha / 2 W'(\langle l_0 \rangle)$ and identifies the MF correlation length along the wedge $\xi_y$ as
\begin{equation}
\xi_y = \left( \frac{\Sigma \langle l_0 \rangle}{W'(\langle l_0 \rangle)} \right)^{\frac{1}{2}}
\label{Wedxiy}
\end{equation}
Substituting for $l_0$ we find the desired MF critical exponent $\nu_y$
 as $\nu_y = 1/p + 1/2$. The important observation here is that $\nu_y \gg \nu_x$ so that the fluctuations are highly anisotropic and dominated
by the modes along the wedge. Indeed as one approaches the bulk phase
boundary $\xi_y$ becomes arbitrarily larger than $\xi_x$ implying that the fluctuations become pseudo-one dimensional. 
  The last lengthscale to calculate within the MF analysis is the midpoint
roughness defined by $\xi_{\perp}^2 \equiv \langle [l(0,y) - \langle l_0
 \rangle]^2 \rangle$ which can be identified from the relation
\begin{equation}
\xi_{\perp}^2 = H(0,0;0)
\label{Wedxiperpdef}
\end{equation}
where the RHS can be obtained from the Fourier inversion of $S(0,0;Q)$. This
leads to the roughness relation for the $d=3$ wedge
\begin{equation}
\xi_{\perp} \sim \left( \frac{\xi_y}{\Sigma l_0} \right)^{\frac{1}{2}}
\label{Wedxiperp}
\end{equation}
which is one of the central results of the MF analysis. Substituting for the
divergence of $\xi_y$ and $l_0$ we find that all dependence on the
binding potential exponent $p$ cancels leading to a remarkable universal divergence
\begin{equation}
\xi_{\perp} \sim t^{-\frac{1}{4}}
\label{Wedxiperpt}
\end{equation}
valid for all ranges of intermolecular potential. In summary, the MF values of
the lenghscale critical exponents at critical filling are  
\begin{equation}
\beta_s^{W} = \frac{1}{p}, \quad \nu_y = \frac{1}{2} + \frac{1}{p}, \quad
\nu_{\perp}^{W} = \frac{1}{4}
\label{WedCoeffs}
\end{equation}
Next we turn to the fluctuation theory of filling beyond MF approximation. 

\subsection{Fluctuation Theory: breather modes and their excitations}

   The MF analysis presented above is only valid if the fluctuations are small
in some sense. Since the divergence of $\xi_{\perp}$, even within MF, is
algebraic we can simply use the `contact condition' \cite{LandF} that
$\xi_{\perp}$ cannot be greater than $\langle l_0 \rangle$ to determine the Ginzburg
criterion for the breakdown of MF theory. Thus we can assert that the above critical exponents are valid provided 
\begin{equation}
\frac{\xi_\perp}{\langle l_0 \rangle} \sim t^{-(\frac{1}{4} -
\frac{1}{p})} \ll 1
\label{Ginz}
\end{equation}
which in the limit $t \rightarrow 0$ implies that $p$ must be less then
$4$. For $p>4$, fluctuation effects dominate and we can anticipate
(analogous to the WFL and SFL regime behaviour of critical wetting) that
the roughness $\xi_{\perp}$ is comparable with the interfacial height 
$\langle l_0 \rangle$. Unfortunately it is very difficult to develop a full fluctuation theory of
filling transitions based on the effective interfacial Hamiltonian \eref{EffHamW}
since, as is clear from the above remarks, the upper critical dimensions for 
wetting and filling are quite different. The situation is rather similar
to that encountered in the theory of wetting where one would ideally
like to develop a fluctuation theory based on analysis of, say, a
semi-infinite Ising or  Landau-Ginzburg Wilson (LGW) model. Such a
approach would afford a description of both bulk (and surface) critical
behaviour and also of wetting. Since this has not proved possible one
instead focuses on effective models constructed to capture the
essential physics of the phase transition in question. Of course, all
Hamiltonians are effective Hamiltonians but some are more effective than
others. The LGW model based on a local magnetization order-parameter can
(in principle) describe bulk and surface criticality in addition to
wetting and filling. In turn the effective interfacial Hamiltonian
\eref{EffHamW} based on a collective co-ordinate $l(x,y)$ is a theory of wetting and
filling. The task here is to develop a simpler model than \eref{EffHamW}
which just affords a description of filling. 

The theory of wedge filling which we propose is constructed from the full
interfacial model under the assumption that the only fluctuations that
determine the asymptotic critical behaviour arise from the pseudo-one
dimensional long-wavelength undulations in the local height of the
filled region. We refer to fluctuations in the local height as breather
modes since these increase/decrease the local width of the filled region
(and the volume of liquid beneath it). Thus we seek to derive a model based
on a collective co-ordinate $l_0 (y)$ which only describes these modes (see
Fig 3.1) and which is valid for sufficiently small wave-vectors $Q \ll
1/\xi_x$. Such a model can be easily constructed by generalising the
method introduced by Fisher and Jin \cite{3DParry, FJ, FJP} for the derivation
of wetting interfacial models from the LGW Hamiltonian. Assuming that
for a given constrained configuration of the mid-point height $l_0 (y)$
that all other fluctuations are small (relative to the soft mode), it follows that the desired filling Hamiltonian can be identified by
\begin{equation}
\exp (-H_W[l_0]) = \mathcal{T} \! \mbox{{\it r}} \left\{ \exp (-H[l]) \right\}
\label{HDef}
\end{equation}
where the $\mathcal{T} \! \mbox{{\it r}}$ denotes a partial trace with respect to the generalised crossing-criterion (GCC) constraint that $l(x = 0,y) = l_0
(y)$. Since this trace is over non-critical fluctuations one can employ a
saddle-point idenifiction
\begin{equation}
H_W[l_0 (y)] = H[l^{\dag}(x,y;l_0 (\cdot))]
\label{Hdef2}
\end{equation}
where $l^{\dag}(x,y;l_0 (\cdot))$ denotes the profile that minimises
\eref{EffHamW} subject to the GCC. As in the Fisher-Jin theory of wetting,
the desired non-planar constrained configurations can be constructed
perturbatively in terms of constrained profiles that are translationally
invariant along the wedge (analogous to the planar constrained
magnetization profiles) which we simply write as $l^{\dag} (x;l_0)$ to
avoid a proliferation of indices. These translationally invariant
profiles satisfy the Euler-Lagrange \eref{E-L} subject to the GCC $l^{\dag}
(x = 0; l_0) = l_0$. The resulting wedge filling model has the form
\begin{equation}
H_W[l_0] = \int dy \left\{ \frac{\sigma(l_0)}{2} \left(\frac{d l_0}{d y}
\right)^2 + V_W (l_0) \right\}
\label{FeffhamDef}
\end{equation}
 where $V_W (l_0)$ denotes the wedge binding potential and $\sigma(l_0)$
 is a position dependent wedge line stiffness. Formally these quantities are determined by the relations
\begin{equation}
V_W (l_0) = \int_{- \infty}^{\infty} dx \left\{ \frac{\Sigma}{2}
\left(\frac{d l^{\dag} (x; l_0)}{dx} \right)^2 + W(l^{\dag}(x; l_0) -
\alpha \vert x \vert) \right\}
\label{Vf}
\end{equation}
(up to unimportant additive constants defined such that $V_W (l_{\pi})=0$) and
\begin{equation}
\sigma(l_0) = \Sigma \int_{- \infty}^{\infty} dx \left( \frac{\partial
l^{\dag} (x; l_0)}{\partial l_0} \right)^2
\label{sigmal}
\end{equation}
  The planar constrained profiles $l^{\dag}(x; l_0)$ have the same simple
near-flat structure as the equilibrium MF profile discussed
earlier. This means that the evaluation of (\ref{Vf}, \ref{sigmal}) is
particularly simple so that for first-order and critical wedge filling we can approximate
\begin{equation}
H_W[l_0] = \int dy \left\{ \frac{\Sigma l_0}{\alpha} \left(\frac{d l_0}{d y}
\right)^2 + V_W (l_0) \right\}
\label{FeffHam}
\end{equation}
  The qualitative and quantitative form of the filling potential depends
  on the order of the MF phase transition. For critical filling it has the expansion
\begin{equation}
V_W (l_0) = \frac{\Sigma (\theta_{\pi}^2 - \alpha^2) l_0}{\alpha} + a_F
l_0^{1 - p} + \dots
\label{FVeff}
\end{equation}
and of course fluctuations are restricted to the region $l_0 > 0$
analogous to the hard-wall condition for wetting. Here the effective
Hamaker constant $a_F \equiv a/(p-1) \alpha$ and note that the
coefficient of the linear term is simply proportional to $t$, the
relevant reduced temperature. The critical filling potential shows a single
minimum (located precisely at the MF value of $l_0$) which continuously
moves out to infinity as $t \rightarrow 0$. The one-dimensional filling
Hamiltonian can be studied using a number of different techniques
including transfer-matrix theory \cite{ourJphysCM, ourPRL2, BandN} and
approximate RG methods \cite{FJP}. However there is no need to go to
these lengths to extract the fluctuation regimes and the values of the
critical exponents since they follow from a simple extension of the
Lipowsky-Fisher scaling theory considered earlier. To see this we
suppose that the influence of fluctuations at critical filling follows
from analysis of an effective potential
\begin{equation}
V_{\mbox{{\it eff}} \; \,} (l_0) \equiv V_W(l_0) + V_{\mbox{{\it fl}} \,} (l_0)
\label{FVeff2}
\end{equation}
which accounts for the entropic repulsion of the unbinding interface from the
wedge bottom. Again the fluctuation term can be estimated from both the
form of the bending energy contribution to \eref{FeffHam} and also the
number of collisions the interface has with the wedge bottom. This we
estimate to of
order $l_0^{3 - 2/ \zeta_{F}}$ where $\zeta_F$ denotes the thermal wandering exponent for wedge
filling (in $d=3$) relating the roughness and pertinent correlation
length $\xi_{\perp} \sim \xi_{y}^{\zeta_F}$. The value $\zeta_F = 1/3$
follows from considering the invariance of the free part of the filling
Hamiltonian under the scale transformation $y \mapsto y/b$ and $l_0
\mapsto b^{-\zeta_F} l_0$. Thus we arrive at an effective filling
potential of the form
\begin{equation}
V_{\mbox{{\it eff}} \; \,} (l_0) \equiv V_W(l_0) + C l_0^{-3}
\label{FVeff3}
\end{equation}
which has a single minimum at the equilibrium mid-point height. In this way we find two
different fluctuation regimes for filling depending on whether the direct
intermolecular potential or entropic contribution determines the
next-to-leading order correction to the linear term in \eref{FVeff}. The
leading linear term is always relevant (in the RG sense) and reflects
the influence of the wedge geometry on the statistical/thermodynamic
properties of filling. This is rather similar to the theory of
fluctuation effects at complete wetting transitions \cite{Comment}. In this
way we predict there exists a filling MF regime (FMF) for $p < 4$, with
exponents given by \eref{WedCoeffs} and a filling fluctuation dominated
regime (FFL) for $p>4$ with
\begin{equation}
\beta_s^{W} = \frac{1}{4}, \quad \nu_y  = \frac{3}{4}, \quad \nu_{\perp}^{W} =
\frac{1}{4} ; \qquad FFL
\label{WedFMF}
\end{equation}
   As a final remark we turn to the effective one-dimensional Hamiltonian 
description of first-order filling. This is based on the model
\eref{FeffHam} but with a different potential $V_W (l_0)$ the form of
which is shown in Fig. 3.3. In fact the structure of $V_W (l_0)$ for
this case can be qualitatively deduced by noticing that the extrema of
$V_W (l_0)$ must correspond to the possible solutions of the MF equation
\eref{E-LCond}. For $\theta_{\pi} > \alpha$, $V_W (l_0)$ has only a single
minimum which remains finite at $T = T_F$. However for $\theta_{\pi} <
\alpha$ a potential barrier appears separating the local minimum and the
infinitely deep potential well at infinity. Note that exactly at the
first-order filling transition temperature the minimum of $V_W (l_0)$
has a finite well-depth relative to $V_W (\infty)$. This is quite unlike
the behaviour of $W(l)$ at first-order wetting transitions. These
remarks indicate even if the mean-field condition for critical filling
is not fulfilled, the influence of the one-dimensional breather mode
fluctuations may still drive the transition critical by causing the
interface to tunnel away from the local minimum of the
potential. Whether this happens or not depends on the range and depth of
the minimum of $V_W (l_0)$ but will
certainly be most important for systems with short-ranged forces (such
as Ising models). Note this tunnelling mechanism does not alter the
filling transition phase boundary \eref{fillT} which simply reflects the
relevance of the linear term in $V_W (l_0)$. One may also note that
this tunnelling must definitely occur out of coexistence for
temperatures $T > T_F$ where it is responsible for the washing out of
the pre-filling line. 

\begin{figure}
\begin{center} \resizebox{0.95\textwidth}{!}{%
  \includegraphics{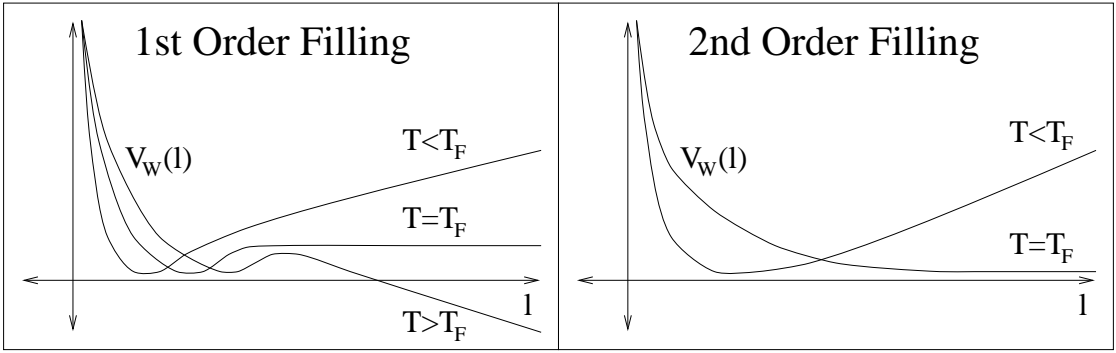}
}
\caption{Schematic illustration of the wedge filling potential on
approaching a first-order and continuous filling transition.}
\end{center}
\label{Fig3}
\end{figure}  

   We mention only briefly here that it is trivial to investigate critical
wedge filling from off-bulk co-existence using the effective model
by allowing for a term of order $l^2$ which couples to the undersaturation in
the filling potential. This clearly represents the contribution from the volume
of undersaturated liquid in the filled region of the wedge. From this one can
readily establish that the filling thickness, roughness and correlation lengths
all exhibit scaling with a gap exponent $\delta = 5/4$ in the FFL regime.Thus
the divergence of the mean interface height along the critical wedge isotherm
as one approaches saturation is characterised by a $-1/5$ power law.

\subsection{Wedge Filling in two dimensions} 

It is a simple matter to extend the wedge filling fluctuation theory to 
arbitrary dimensions $d$ and in particular $d=2$ where one can compare the 
predictions with the exact transfer matrix solution of the full interfacial 
model \eref{EffHamW}. Consider a wedge in bulk $d$ dimensions which is of `V' shape cross-section in
the $x$ direction but which is translationally invariant in
the remaining $d-2$ (surface) dimensions. Denoting the position of the
interface above the wedge bottom by $l(\vec{x}_{\parallel})$ where
$\vec{x}_{\parallel}$ denotes the $(d-2)$ dimensional vector parallel to
the wedge orientation. The filling Hamiltonian trivially generalises to
\begin{equation}
H_W[l_0] = \int d \vec{x}_{\parallel} \left\{ \frac{\Sigma l_0}{\alpha}
\left(\nabla l_0 \right)^2 + V_F (l_0) \right\}
\label{Effham2D}
\end{equation}
which clearly describes the $(d-2)$ dimensional thermal excitations of 
the breather interfacial mode. The scaling theory also generalises in a natural
way. First we calculate the $d$ dimensional wedge wandering exponent
$\zeta_W$ by considering the invariance of the free part of \eref{Effham2D}
under scale transformations. This determines $\zeta_W$ as      
\begin{equation}
\zeta_W = \frac{4 -d}{3}
\label{zetaWed}
\end{equation}
relating the roughness and correlation length measured along the wedge
(which we here denote $\xi_{y}$) via $\xi_{\perp} \sim \xi_{y}^{\zeta_W}$. Similarly the fluctuation
contribution to the effective potential arising from the bending energy and entropy terms scales as
\begin{equation}
V_F (l_0) \sim l_0^{-\frac{3(2-d)}{4 - d}} \equiv l_0^{-\tau_W}
\label{FVeff2D}
\end{equation}
where $\tau_W$ is the wedge fluctuation exponent analogous to $\tau$ in the Lipowsky-Fisher theory of wetting. 
  
This heuristic scaling theory predicts that for general dimension $d <
4$, the criticality at filling falls into a mean-field (FMF) and
fluctuation dominated (FFL) regime depending on whether $p$ is less or
greater than the marginal value $p^* = 2(d-1)/(4 - d)$. The critical exponents are given by
\begin{equation}
\beta_s^{W} = \frac{1}{p}, \quad  \nu_y = \frac{1}{p} + \frac{1}{2},
 \quad \nu_{\perp}^{W} = \frac{4 - d}{4} + \frac{(3-d)}{2p} ; \qquad FMF
\label{2DCosFMF}
\end{equation}
\begin{equation}
\beta_s^{W} = \nu_{\perp}^{W} = \frac{(4 - d)}{2(d - 1)}, \quad  \nu_y =
\frac{3}{2(d-1)} ; \qquad FFL
\label{2DCosFFL}
\end{equation}
and notice that it is only in $d=3$ that the roughness exponent is
universal independent of the range of the forces. These predictions for
critical singularities at wedge filling are, in general very different to
the predictions for critical wetting. However for the special case of
$d=2$ there appears to be a connection between filling and critical wetting discussed below. In $d=2$ the breather fluctuation theory predicts that the marginal intermolecular range is $p=1$ and that the values of the FFL regime critical exponents are
\begin{equation}
\beta_s^{W} = \nu_{\perp}^{W} = 1; \qquad d = 2 \; \; FFL
\label{2DCosSFL}
\end{equation}
with $\nu_y$ undefined. Note that these are pertinent to all physical
examples of intermolecular forces. The first thing to observe here is that
the numerical values of the filling critical exponents $\beta_s^{W}$ and
$\nu_{\perp}^{W}$ in the $d=2$ FFL are identical to the analogous critical wetting exponents in the ($d=2$) SFL regime. As we shall now see this coincidence extends to the full form of the PDF. 
   
As well as predicting the critical exponents at filling the fluctuation
theory also predicts the position dependence of the scaling form of the
mid-point PDF, $P_W (l_0)$. To see this note that the case of wedge
filling in $d=2$ is special because the breather mode is only subject to
zero dimensional fluctuations since there is no direction along the
wedge. As a result one can go beyond the above predictions for the
values of the critical exponents and assert that in the asymptotic
critical region, the mid-point PDF is given by
\begin{equation}
P_{W} (l_0) = N e^{-V_F (l_0)} 
\label{2dPDF1}
\end{equation}
where here (and below) $N$ denotes a suitable normalization factor. Now from 
\eref{FVeff} we have $V_F (l_0) \sim \Sigma (\theta_{\pi}^2 - \alpha^2)
l_0 / \alpha \approx 2 \Sigma (\theta_{\pi} - \alpha) l_0$ so that in the critical regime the breather fluctuation theory leads to the scaling form of the PDF
\begin{equation}
P_{W} (l_0) = N e^{-2 \Sigma (\theta_{\pi} - \alpha) l_0} 
\label{2dPDF2}
\end{equation}
or equivalently
\begin{equation}
P_{W} (l_0) = \frac{1}{\langle l_0 \rangle} e^{-\frac{l}{\langle l_0 \rangle}}
\label{2dPDF3}
\end{equation}
which is of course consistent with the values of the critical exponents quoted
in \eref{2DCosFFL}. 
   This prediction is remarkable for two reasons:
(I) it is in exact agreement with the transfer matrix result for the PDF based
on the ($d=2$) interfacial model \eref{EffHamW}. For quite arbitrary choices of
binding potential the mid-point interface height distribution function 
is given by \cite{ourPRL1, ourJphysCM}
\begin{equation}
P_W (l_0) = N e^{2 \Sigma \alpha l_0} \vert \Psi_0^{(\pi)} (l_0) \vert^2
\label{2dPDF4}
\end{equation}
where $\Psi_0^{(\pi)} (l)$ is the ground state eigenfunction appearing
in the transfer matrix spectrum for $d=2$ wetting at a planar wall. This
satisfies the well known Schr\"{o}dinger equation \cite{Burkhardt}
\begin{equation}
-\frac{1}{2 \Sigma} \frac{\partial^2}{\partial l^2} \Psi_0^{(\pi)} (l) + W(l) \Psi_0^{(\pi)} (l) = E_0
\Psi_0^{(\pi)} (l)
\label{Schr}
\end{equation}
where $E_0$ is the corresponding groundstate eigenvalue which is related
to the contact angle by $E_0 = -\Sigma \theta_{\pi}^2 / 2$. From the
properties of $\Psi_0^{(\pi)} (l)$ it follows that the filling transition is indeed characterised by the scaling regimes and corresponding critical singularities predicted by the general fluctuation theory. Moreover
for $p > 1$,  and far from the wall the wave function decays as 
\begin{equation}
\Psi_0^{(\pi)} (l) \sim e^{- \Sigma \theta_{\pi} l}
\label{Wavefn}
\end{equation}
so that \eref{2dPDF2} and \eref{2dPDF4} are identical. The equivalence
of the scaling form of the PDF with the breather mode fluctuation theory
and the full effective interfacial model completely supports our
conjecture that it is the overall breather mode which translates the
interface up and down the sides of the wedge, rather than the thermal
wandering of the interface within the filled region that determines the
criticality at filling. 

(II) The derived expression for the filling PDF in the FFL regime has an
identical structure to the analogous interfacial height PDF, $P_{\pi}
(l)$ for critical wetting in the SFL. Recall that this quantity is
determined as $P_{\pi}(l) \sim \vert \Psi_0^{(\pi)} (l) \vert^2$ with
the transfer matrix approach so that we in the SFL we have the well
known result, quoted earlier
\begin{equation}
P_{\pi}(l) = \frac{1}{l_{\pi}} e^{-\frac{l}{l_{\pi}}}
\label{2dPDF5}
\end{equation}
which is identical to the result for filling. In fact in
$d=2$ the connection between filling and critical wetting for purely
thermal disorder extends to the case marginal case $p =1$ for filling
where the structure of the FMF/FFL PDF has the same form as the PDF for
critical wetting at a WFL/MF borderline for which the specific heat
exponent vanishes \cite{ourJphysCM}. The PDF for this case is not simply a pure exponential function. 
  
The equivalence of the film thickness (and roughness) critical exponents and PDF's for two dimensional, fluctuation-dominated filling and wetting has also been established for the case of filling/wetting with random bond disorder described by the interfacial model \cite{ourJphysCM}
\begin{equation}
H[l] = \int \left\{ \frac{\Sigma}{2}(\frac{\partial l}{\partial
x})^2+V_r(x,l(x))+ W(l-\alpha \vert x \vert) \right\}
\label{RandH}
\end{equation}
with disorder averages satisfying
\begin{equation}
\overline{V}(x,l(x)) = 0
\label{av1}
\end{equation}
and
\begin{equation}
\overline{V_r(x,l(x))V_r(x',l'(x))} = \Delta \delta(x-x') \delta (l(x)-l'(x'))
\label{av2}
\end{equation}
where $\Delta$ is a measure of the disorder. The study of filling transitions in this model allows us to test the
possible connection between fluctuation dominated two dimensional filling and critical wetting in a system where the interfacial wandering exponent is different to the thermal case. It transpires that for systems with purely short-ranged forces the model
\eref{RandH} can be solved exactly using the transfer 
matrix method by extending Kardar's Bethe Ansatz approach \cite{Kardar}
used in his analysis of wetting. This again confirms the
thermodynamic prediction for the location of the filling transition and
identifies the universal filling critical exponents as
\begin{equation}
\beta_s^{W} = 2, \quad \nu_{\perp}^{W} = 2
\label{RandCos}
\end{equation}
which are identical to the analogous critical exponents $\beta_s$ and
$\nu_{\perp}$, determined by Kardar for critical wetting with random
bonds with short-ranged forces corresponding to the SFL regime
\cite{Forgacs, Kardar, Fisher}. The expression for the scaling form of
the mid-point interfacial height PDF $P_F (l_0)$ for $d=2$ random bond filling is highly
non-trivial but again analysis shows that it is identical to the PDF for random
bond wetting \cite{ourJphysCM}. 

\section{Conic Filling.}

\subsection{Phenomenology.}
  
Consider a non-planar wall-fluid interface in the shape of an
infinite cone which makes an angle $\alpha$ to the horizontal. Thus, for open
wedges (small $\alpha$) the height
of the wall above the plane is described by a height function $z_C
(\vec{r})=\alpha \vec{r}$ with $\vec{r}$ the displacement vector along the
two-dimensional displacement vector along the horizontal plane with origin at
the cone apex. As before the 
wall is supposed to be in contact with a
vapour phase at two-phase coexistence with thermodynamics indicating that
the cone is completely filled for $T > T_F$, where the conic filling
temperature $T_F$ satisfies the same condition $\theta_{\pi} (T_F)=
\alpha$ as filling in the $d=3$ wedge \eref{fillT}. We allow for the
possibility of first-order and critical cone filling corresponding to the
discontinuous or continuous divergence of the mid-point filling height
$\langle l_0
\rangle $ as $t \equiv (T_F - T)/T_F \rightarrow 0$. Also of interest is the roughness
$\xi_{\perp}$ arising from the interfacial fluctuations of the unbinding
liquid vapour interface. The radius of the near flat, filled region of
the cone, which also characterises height fluctuations in the central
region is trivially related to the height via $\xi_r \sim \langle l_0
\rangle / \alpha$. As with wedge filling we anticipate that for radial positions
$r \gtrsim \xi_r$ the interface will simply take the microscopic value
$l_{\pi}$ it would have on a planar wall. Thus we define two lengthscale critical exponents for critical cone filling by
\begin{equation}
\langle l_0 \rangle \sim t^{-\beta_s^{C}}, \quad \xi_{\perp} \sim t^{-\nu_{\perp}^{C}}
\label{coneCos}
\end{equation}
in an obvious notation. Here we seek to understand both the conditions
under which critical cone filling can occur and the values of the
critical exponents at MF level and beyond. As we shall show these are
quite distinct from the situation for wedge filling. 
      
Following our earlier treatment of wedge filling we anticipate that
interfacial fluctuations manifest themselves in two different ways at
conic filling. Firstly there is the usual wandering of the interface
within the filled region determined by the value of the standard
wandering exponent $\zeta_{\pi}$ which is of course marginal in $d=3$
(concentrating on thermal fluctuations). Such fluctuations would occur
even if the edges of the filled region were fixed to a given height,
$l_0$ say, and
would contribute terms of order $\sqrt{l_0}$ and $\sqrt{ \ln l_0}$ to the
interfacial roughness in $d=2$ and $d=3$ respectively. However much more
important than this is the breather mode that translates the entire flat
region up and down the cone. The fundamental distinction between
three-dimensional wedge
and cone filling arises because the breather mode is
thermally excited by quasi-one dimensional and zero-dimensional
excitations respectively. In this respect three dimensional conic
filling is the higher dimensional version of two-dimensional filling
considered in the last section. Note that the effective dimensionality of the breather mode does not destroy the phase transition because the fluctuating field diverges as the filling temperature is approached.

\subsection{Mean-field theory. }
 
Denoting the local interfacial height above the plane by $l(\vec{r})$, the
natural starting point for the study of conic filling for open cones
(small $\alpha$) is the standard interfacial model
\begin{equation}
H[l] = \int d \vec{r} \left\{ \frac{\Sigma}{2} \left(\nabla l
\right)^2 + W(l - \alpha {r}) \right\}
\label{ConeEffHam}
\end{equation}
where, again, $\vec{r}$ denotes the radial vector from the apex. To begin we
chose binding potentials $W(l)$ of the type \eref{BPot} corresponding
to walls that exhibit critical wetting and set the exponent parameter $q
= p - 1$. Numerical minimization of the model
Hamiltonian for different ranges of the intermolecular potential shows that
the filling transition is critical (for the present choice of critical wetting
walls) and located precisely at $\theta_{\pi} = \alpha$. However the
numerics also indicates that the MF value of the critical exponent
characterising the divergence of $\langle l_0 \rangle$ appears to fall into two distinct
regimes. For $p > 1$ the critical behaviour is universal
and dominated by the geometry such that $\beta_s^{C} = 1$. On the other
hand if $p < 1$ the critical exponent is consistent with the
identification $\beta_s^{C} = 1/p$, identical to that found within the MF
theory of wedge filling. This intriguing MF critical behaviour follows
from a very simple and highly accurate variational theory of conic
filling. Noting the simplicity of the equilibrium interfacial profiles
obtained in our numerical minimization we parameterise the possible height function by
\begin{equation}
l^{\dag} (\vec{r}) = \left \{ \begin{array}{ll}
l_0, &  {r} < \frac{l_0 - l_{\pi}}{\alpha} \\
l_{\pi} + \alpha {r}, & {r} > \frac{l_0 - l_{\pi}}{\alpha} 
\end{array}
\right.
\label{ldag}
\end{equation}
and determine the mid-point height dependence of the conic filling potential
defined by
\begin{equation}
V_c (l_0) = H[l^{\dag}] - H[l^{\dag}] \Big|_{l^{\dag} (0) = l_{\pi}}
\label{coneVeff}
\end{equation}

  Note that this variational approximation becomes increasingly accurate for
as one approaces a critical filling transition since the profile tends to a
macroscopically flat meniscus. Minimization of the conic filling potential leads to the following
  integral equation for the mid-point height
\begin{equation}
\frac{\Sigma (\alpha^2 - \theta_{\pi}^2) (l_0 - l_{\pi})}{2} = \int_{l_{\pi}}^{l_0} W(l) dl
\label{Conemaster}
\end{equation}

This equation has an simple graphical interpretation which recovers the
numerical results quoted above and leads to an elegant criterion for the
order of the filling transition. The filling transition is continuous provided
\begin{equation}
\int_{l_{\pi}}^{\infty} W(l) dl < 0
\label{ConeCond}
\end{equation}
corresponding to an integral measure of the overall attractive/repulsive
nature of the binding potential. By allowing for pertubations to the simple
variational ansatz it is easy to show that this is the exact MF condition for 
criticality
within the effective interfacial model to order $\mathcal{O} (\alpha^4)$. The condition for criticality of conic
filling is much less severe than that for both critical wetting and
critical wedge filling and implies that even substrates which exhibit
rather strong first order-wetting will exhibit a continuous filling
transition in a cone geometry. A comparison of the respective shapes of
the binding potential need for criticality for three dimensional
wetting, wedge filling and cone filling is shown in Fig. 4.1. Provided
the condition for critical cone filling is met the MF critical exponent
falls into two regimes:
\begin{equation}
\beta_s^{C} = \frac{1}{p} \mbox{  for } p < 1 \mbox{ and } \beta_s^{C} = 1 \mbox{ 
for } p > 1
\label{ConeCos1}
\end{equation} 

\begin{figure}
\begin{center}\resizebox{0.6\textwidth}{!}{%
  \includegraphics{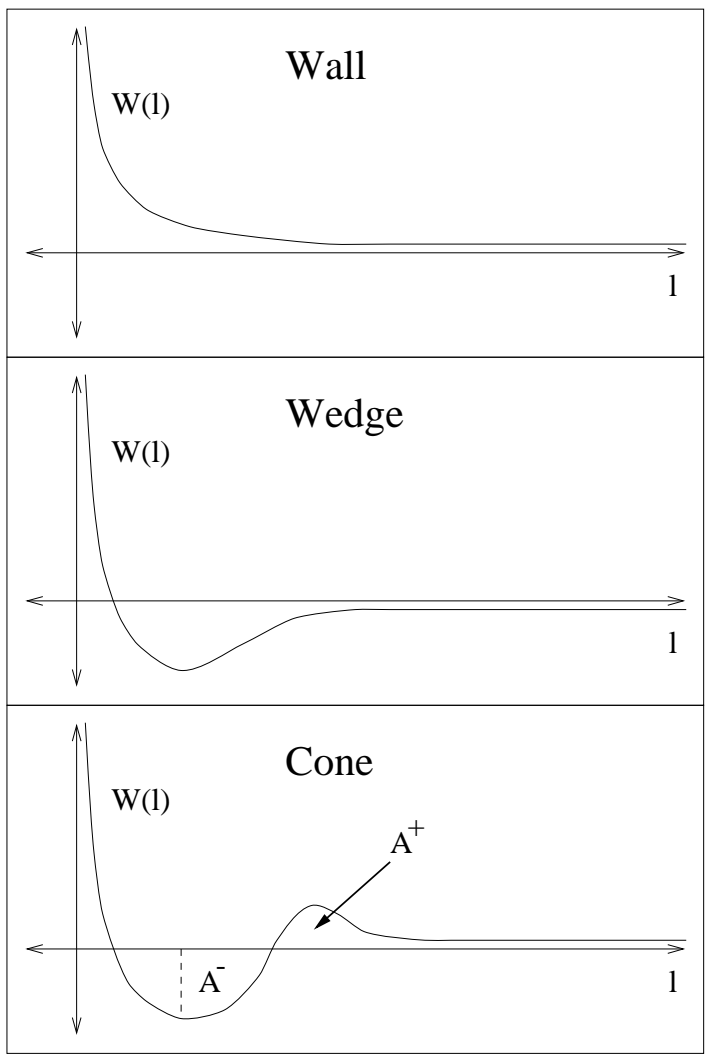}
}
\caption{Schematic illustration of the qualitive form of the binding
potential $W(l)$ necessary for a continuous wetting/filling
transition. For critical wetting at a planar wall (top) the potential
decays monotonically exactly at $T_{\pi}$. For a wedge, the potential
must show no local maximum at $T_F$. For a cone the integrated stengh of
the potential must be attractive ($A^{+} < A^{-}$).}
\end{center}
\label{Fig4}
\end{figure}  

The universality of the critical behaviour in the geometry dominated
critical regime is reminiscient of the way in which fluctuation effects
lead to universal critical behaviour at wetting transitions. This
connection is explored further below.
\subsection{Fluctuation theory: breather modes and the PDF.}

The effective zero-dimensionality of the breather mode at conic filling
suggests that we should analyse fluctuation effects in the same manner as we
did for filling of a wedge/cone in $d=2$. Recall that this simple approach
gave the precise scaling form of the mid-point interfacial height PDF
(and hence the critical exponents) at filling as that found from the
exact transfer-matrix theory. Accordingly we propose that the critical
behaviour at conic filling is described by a PDF for the mid-point
interfacial height is of the form
\begin{equation}
P_C (l_0) = N e^{-V_C (l_0)}
\label{conePDF}
\end{equation} 
since the role of the cone potential in $d=3$ is analogous to the wedge
potential $V_W (l_0)$ in $d=2$. It is a simple matter to show that $V_c
(l_0)$ has the
form 
\begin{equation}
V_C (l) = \frac{\Sigma (\theta_{\pi}^2 - \alpha^2) l^2}{\alpha} - \sigma_C
l + a_C l_0^{2 - p} + \dots
\label{coneVeff2}
\end{equation}
consistent with the MF values of the critical exponents discussed
above. The three terms in this expression represent the purely geometrical 
contributions arising form the area and circumference of the filled region and also the direct influence of the specific 
intermolecular forces. Both $\sigma_C$ and $a_c$ are positive effective Hamaker
constants. For $p > 1$ the final term in \eref{conePDF} is irrelevant, geometry dominates and the PDF can be written
\begin{equation}
P_C(l_0) = N e^{-\frac{1}{2 \xi_{\perp}^2} (l_0 - \langle l_0 \rangle)^2}
\label{conePDF2}
\end{equation}
and identifies the universal values of the critical exponents as
\begin{equation}
\beta_s^C = 1 \mbox{ and } \nu_{\perp}^{C} = \frac{1}{2} \mbox{ for }  p > 1
\label{conecos}
\end{equation}
On the other hand if the second term in \eref{coneVeff2} is irrelevant
and close to its most probable position the PDF is Gaussian with critical exponents given by
\begin{equation}
\beta_s^C = \frac{1}{p} \mbox{ and } \nu_{\perp}^{C} = \frac{1}{2} \mbox{ for } p < 1
\label{conecos2}
\end{equation}
  Equations \eref{ConeCond} and (\ref{conePDF}-\ref{conecos2})
  are the main predictions of this section. Here we make two pertinent remarks:

(I) the MF values of the filling height critical exponent $\beta_s^{C}$ is
unchanged by fluctuations. However the thermal fluctuations of the
breather mode do lead to an appreciable interfacial roughness. The
predictions \eref{conecos} are appropriate for both experimental systems
with long-ranged dispersion forces and simulation studies of Ising-like
systems with short-ranged forces. For discrete lattice based Ising
studies, it is probably more convenient to study adsorption in an
inverted pyramid shaped geometry. The critical exponents are the same as
the cone for this case. 

(II) the scaling form of the PDF and the values of the critical
exponents in the universal geometry dominated regime of cone filling are
identical to the analogous critical singularities characteristic of the
SFL regime of critical wetting short-ranged forces i.e. regime (III)
corresponding to $\omega > 2$. This is directly
analogous to the observations made earlier for wedge/cone filling in $d=2$. 

The above predictions point to a remarkable coincidence between the
universality of conic filling in $d=2$ and $d=3$ and the SFL regime of critical
wetting in the same dimension which is encapsulated in the equivalence of the
film thickness and roughness critical exponents as well as the interfacial 
height PDF's. Thus in the important geometry dominated regimes of cone
filling and for the physically relevant dimensions $d = 2$ and $d = 3$
one can identify
\begin{equation}
\beta_s^C = \beta_s^{SFL}, \quad \nu_{\perp}^{C} = \nu_{\perp}^{SFL} \mbox{ and }
P_C(l_0) = P_{\pi}^{SFL} (l_0)
\end{equation}

\subsection{Universal conic filling and the SFL regime. }

The above predictions beg the question, why does the influence of the
cone geometry on the (universal) unbinding at filling mimic the
fluctuations characteristic of critical wetting in the SFL? Usually, of
course, if two different systems exhibit the same critical exponents one
concludes, after identifying the appropriate order-parameter, that they
belong to the same universality class. We do not believe that this is
the correct explanation because in this case there does not appear to be an analogue
of the wetting correlation length critical exponent $\nu_{\parallel}$
for filling. In other words the equivalence of the one-point PDF's does
not extend to the two point function. In view of this we forward a
simple plausibility argument suggesting that the equivalence of the PDF's
is not simply coincidence but {\it is} special to the physical dimensions
$d=2$ and $d=3$. We emphasise that the specific results of our model
calculations for the values of the critical exponents stand quite apart
from this discussion. 
   
Let us focus on the structure of the PDF for critical wetting at a planar
wall in $d=2$ and $d=3$. The distribution function $P_{\pi} (l)$ is found by
evaluating the (normalised) Boltzmann sum over all interfacial
configurations that pass through the specified height $l$ at an arbitarily chosen position
$\vec{r}=0$ (say),
equivalent to the identification $P_{\pi} (l) = N e^{-F^{*}[l]}$ where
$F^{*} = - \ln Z_1^*$  and $Z_1^{*} $ is the partial partition function
summing over this class of configurations. In turn the constrained
free-energy $F^{*} [l]$ is a functional of the constrained height profile
corresponding to the (partial) average over all profiles that satisfy
the appropriate condition $l(x = 0) = l_0$. For large $l \gg l_{\pi}$
the constrained profile determining the PDF naturally forms the shape of
a triangle (in $d=2$) and cone (in $d=3$) characterised by the equilibrium contact angle
$\theta_{\pi}$. Thus the
droplet shape of the constrained profile determining the asymptotics of $P_{\pi}(l)$
near critical wetting is essentially identical to the equilibrium height
profile measured with respect to
the local wall height near critical cone filling (since $\theta_{\pi}
\sim \alpha$). This observation hints at a possible connection between
the PDF's at wetting and cone filling. One may also reason that if there
is any connection between the PDF's of filling and wetting then it will
only be valid when the direct effect of intermolecular interactions is
unimportant since the non-universal influence of these forces manifests itself in very different ways. This is only true in the geometry dominated regime of cone filling and the SFL regime of critical wetting. 
  
On the other hand it is easy to see that any connection between the two
PDF's is not valid for all dimensions. The generalisation of our simple
fluctuation theory of cone filling presented to arbitary dimension $d$
is immediate. Using the same variational approach described above we
arrive at a cone binding potential of the form (neglecting unimportant
constants)
\begin{equation}
V_C(l) = t l^{d -1} - l^{d-2} + \cdots
\label{conePDF3}
\end{equation}
representing the leading order surface and line contributions. The influence of
intermolecular forces is present through a term $\mathcal{O} (l^{d - 1 -
p})$ which is higher-order for $p>1$ independent of the
dimensionality. Using this cone potential to evaluate the PDF arising
from the (zero-deimensional) thermal fluctuations of the breather mode leads to the
following prediction for the universal, geometry dominated cone filling
critical singularities in arbitrary dimension $2 <d < 4$,     
\begin{equation}
\beta_s^{C} = 1 \mbox{  and  } \nu_{\perp}^{C} = \frac{4 - d}{2}
\label{coneCos2}
\end{equation}
which of course recovers our results for wedge/cone filling in $d=2$ and cone
filling in $d=3$ as special cases. For $d > 4$ the roughness remains
finite. Only for the physically important dimensions $d=2,3$ to the numerical
values of these critical exponents coincide with the predictions for SFL regime
critical wetting. Nevertheless we find it remarkable that for these special
dimensions not only the values of the critical exponents but also the detailed
form of the PDF's are identical for filling and critical wetting. Finally we
mention that the exponents quoted above allow us to identify the value of the
wandering exponent for critical cone filling defined by $\xi_{\perp} \sim
\xi_{r}^{\zeta_C}$. Noting that interfacial height is trivially related to the
radial correlation length (width) of the filled region we are led to the
identification 
       
\begin{equation}
\zeta_C = \frac{4 - d}{2}
\label{conedelta}
\end{equation}
which shows the influence of the zero-dimensional breather mode on the
distribution of matter. Obviuously in two dimensions $\zeta_C = \zeta_W = 1$
since the cone and wedge geometries are identical.

\section{Conclusions. }

In this article we have presented details of a general fluctuation theory of
filling transitions at wedges and cones. The main results and
conclusions of our study are summarised below:

(I) The dominant interfacial fluctuations at wedge and cone filling correspond
to  breather-like motion that translates the flat, filled region up and down
the sides of the confining geometry. 

(II) For three dimensional wedge filling the thermal excitations of the
breather mode are described by a novel one-dimensional effective Hamiltonian
which predicts a universal value of  $1/4$ for the roughness critical exponent for all
ranges of the intermolecular forces and also large scale interfacial roughness 
and universal critical singularities for $p > 4$. 

(III) For filling in $d=2$, the simple picture of an effective
zero-dimensional fluctuating breather mode correctly captures the
critical exponents and precise scaling behaviour of the mid-point height
PDF as found from the full transfer-matrix solution to the full interfacial model (of wetting and filling). 

(IV) For two-dimensional filling with both thermal and random bond disorder the
scaling of the mid-point height PDF is identical to the corresponding results
for SFL regime critical wetting in this dimensionality. 

(V) Conic filling shows two different types of critical
behaviour including a universal geometry dominated regime for $p > 1$
pertinent to all physical types of intermolecular force. Applying the
simple breather-like fluctuation picture to this phase transition yields
a scaling form for the PDF and associated interfacial height and
roughness critical exponents identical to the SFL regime prediction for
critical wetting in this dimensionality. We have argued that this equivalence 
of the PDF's is special to $d=2$ and $d=3$.

(VI) Even at MF level the conditions under which three dimensional wedge
and cone filling are continuous are different to the fine tuning
required for critical wetting and are sensitive to the particular
geometry. For critical wedge wetting one requires that the binding
potential shape at $T = T_F$ shows no local maximum. On the other hand
cone-filling is critical provided the integrated weight of the potential
remains negative (attractive). 

These conclusions, regarding both the conditions for criticality and the
manifestations of interfacial fluctuations at wedge and cone filling, are
in sharp contrast to the situation for critical wetting. Concentrating
on the values of the thickness and roughness critical exponents as
pertinent to realistic three-dimensional systems with non-retarded van
der Waals forces ($p=2$) we have 
\begin{equation}
\beta_s = 1 \mbox{ and } \nu_{\perp} = 0 \; (\ln) \mbox{ for critical wetting}
\end{equation}
\begin{equation}
\beta_s^W = \frac{1}{2} \mbox{ and } \nu_{\perp}^W = \frac{1}{4} \mbox{
for critical wedge filling}
\end{equation}
\begin{equation}
\beta_s^C = 1 \mbox{ and } \nu_{\perp}^C = \frac{1}{2} \mbox{ for critical
cone filling}
\end{equation}
indicating that the geometry has a profound influence on the roughness
of the interface. This reflects the special role of the breather mode at
filling transitions. Consequently if critical wetting at a wall-fluid
interface were ever seen experimentally, the observation of wedge and
cone filling in this system would certainly provide a means of observing
large scale interfacial fluctuations. It should certainly be possible to
test the predictions of universal and large scale interfacial
fluctuation related behaviour in Ising model systems.   
   We have also emphasised that the geometry dramatically alters the
conditions for criticality and certainly relaxes the requirement of fine
tuning the values of the Hamaker constants necessary for critical
wetting. The condition for inducing continuous filling transitions is
easiest to fulfil for the cone geometry and is highly encouraging that
the observation of critical filling in cones made from walls exhibiting
first-order wetting is experimentally feasible. It may even be possible
to further manipulate the conditions for criticality by considering
filling in other geometries and by introducing chemical heterogeniety. 
  Throughout this article we have focused exclusively on the filling
transition occuring as $T \rightarrow T_F$ at bulk two-phase
coexistence. Also of interest is the influence of the geometry on
complete wetting adsorption isotherms corresponding to the continuous
divergence of the filling thickness for temperatures $T > T_F$ as $\mu
\rightarrow \mu_{\mbox{sat}}$ \cite{Hauge, Carlos1, Carlos2}. These transitions have also
received much recent attention and show great sensitivity to the shape
of the confining geometry. For cones and wedges in all dimensions the
divergence of the filling height is super-universal and given by
\begin{equation}
l_0 \sim \frac{1}{\delta \mu} \mbox{ and } \xi_{\perp} \sim \frac{1}{\delta \mu^{\zeta_{\pi}}}
\end{equation}
where $\delta \mu$ denotes the undersaturation. However unlike critical filling
the breather mode does not change the nature of the roughness so that
$\xi_{\perp} \sim \delta \mu^{\zeta_{\pi}}$.
 It would be very interesting to explore the possible connection between conic
filling and SFL critical wetting in two and three dimensions in more detail. 
Whilst for $d=2$ the equivalence of the
critical exponents and PDF's has been establised exactly using transfer-matrix
methods it would be extremely instructive to understand this from a RG
perspective. This would perhaps throw light on a more detailed theory of three
dimensional cone filling based on the full interfacial model. Also our analysis
of wedge and conic filling in $d=3$ was limited to pure thermal disorder
unlike our analysis in $d=2$. Studies of three dimensional wedge and
cone filling in impure systems would be very illuminating. It may be
that for impure systems, filling and SFL critical wetting only share common features in $d=2$.
  
AJW and CR acknowledge support from the EPSRC and the E.C.\ under
contract ERBFMBICT983229 respectively.


\begin{thebibliography}{30}
\bibitem{Comment} The linear RG theory of wetting (see for example [3])
     can be applied to the wedge filling Hamiltonian in the usual way by
     treating the potential term pertubatively $W(l)$ and expanding about the
     flow of the free contribution to $H_F [l_0]$ which is approximately
     invariant under rescaling with the filling wandering exponent
     $\zeta_F = 1/3$. In contrast to wetting the RG scheme is no longer
     exact at first-order because of the mixing of slow and fast modes
     arising from the position dependence of the effective line
     tension. The linear term in filling potential is always relevant
     and the approximate linear scheme should correctly describe the
     critical behaviour analogous to the RG theory of complete wetting.
\end{thebibliography}
\end{document}